\documentstyle[aps,pre,psfig,floats]{revtex}

\begin{document}

\twocolumn[\hsize\textwidth\columnwidth\hsize\csname
@twocolumnfalse\endcsname

\title{Traveling wave solutions in the Burridge-Knopoff model}

\author{C. B. Muratov}

\address{Courant Institute of Mathematical Sciences, New York
  University, \\ 251 Mercer St., New York, NY 10012}

\date{\today}

\draft

\maketitle

\begin{abstract}

The slider-block Burridge-Knopoff model with the Coulomb friction law
is studied as an excitable medium. It is shown that in the continuum
limit the system admits solutions in the form of the self-sustained
shock waves traveling with constant speed which depends only on the
amount of the accumulated stress in front of the wave. For a wide
class of initial conditions the behavior of the system is determined
by these shock waves and the dynamics of the system can be expressed
in terms of their motion. The solutions in the form of the periodic
wave trains and sources of counter-propagating waves are analyzed. It
is argued that depending on the initial conditions the system will
either tend to synchronize or exhibit chaotic spatiotemporal behavior.

\end{abstract}


\vskip2pc]

\bibliographystyle{prsty}

\section{Introduction}

Propagating self-sustained waves (autowaves) and more complex
spatiotemporal patterns are characteristic of excitable media of
different nature. A typical example of such a phenomenon is burning of
black powder in a safety fuse. When the fuse is ignited at one end,
the exothermic reaction releases the heat which is then spread out by
heat diffusion. Thus, the neighboring regions of the fuse ignite
leading to self-sustained propagation of the combustion front. The
phenomenon of non-attenuated propagation of waves is in fact common
for a variety of physical, chemical, and biological systems
\cite{cross93,vasiliev,mikhailov,field,murray,ko:book}. Traveling
waves are experimentally observed in semiconductor and gas plasma,
semiconductor and superconductor structures, combustion systems,
active optical media, magnetic media under illumination, autocatalytic
chemical reactions, nerve and heart tissue (see
\cite{cross93,vasiliev,mikhailov,field,murray,ko:book} and references
therein).

In order for self-sustained waves to be feasible, the system must
possess two basic ingredients. First, the system must be {\em
excitable}, that is, there has to be a threshold below which the
perturbation of the steady homogeneous state of the system decays,
while perturbations of larger amplitude grow. In the example above, a
sufficient amount of heat is needed to ignite black powder. Second,
there has to be coupling between the regions of the system at
different points in space. In the case of the safety fuse such a
coupling is provided by heat diffusion leading to spread of the
temperature and ignition of powder in front of the combustion
zone. Thus, the prototype systems exhibiting self-sustained waves are
reaction-diffusion systems
\cite{cross93,vasiliev,mikhailov,field,murray,ko:book}.

Recently, it was pointed out that an entirely different class of
systems may be considered as excitable \cite{cartwright97}. These are
elastic media with friction exhibiting stick-slip motion. Both
experimental observations and numerical simulations show that such
systems are capable of supporting steadily propagating solitary waves
in the form of shocks
\cite{cartwright97,rubio94,schmittbuhl93,espanol94}. These systems are
also of special interest because they are used for modeling the
dynamics of earthquakes
\cite{cartwright97,rubio94,schmittbuhl93,espanol94,%
burridge67,carlson89,carlson91,carlson94,langer91,myers93,xu94}. It is
clear that systems exhibiting stick-slip motion have both necessary
ingredients of excitable systems. The threshold behavior here is due
to static friction, which prevents any motion in the system until some
critical amount of stress is accumulated. The coupling of the elements
of the system at different points in space is due to the non-locality
of elastic stress.

Singular perturbation techniques proved to be very effective in
treating problems of traveling wave propagation in reaction-diffusion
systems
\cite{cross93,vasiliev,mikhailov,field,murray,ko:book,fife,mo1:98}.
These methods use strong separation of time scales in the problem to
decompose the dynamics of the system into fast and slow
motion. Clearly, this situation is also realized in the models of
stick-slip motion where (especially in the context of earthquakes)
there is a strong separation of time scales between fast slipping
events and slow accumulation of stress. It is therefore advantageous
to try to apply these techniques to the problem of stick-slip motion.

In this paper we present a study of the Burridge-Knopoff slider-block
model \cite{burridge67} with the Coulomb friction law. We will show
that for sufficiently slowly spatially varying displacement variable
the dynamics of the system is dominated by self-sustained traveling
shock waves. We will study the properties of these waves and
reformulate the dynamics of the system in terms of their motion.

Our paper is organized as follows. In Sec. II we introduce the
governing equations for the model we study and discuss the features of
the friction law used, in Sec. III we construct the solutions in the
form of self-sustained traveling shock waves, in Sec. IV we
reformulate the dynamics of the system in terms of the motion of these
shock waves and study general properties of the reduced problem, in
Sec. V we analyze two different types of solutions, and in Sec. VI we
draw conclusions.

\section{model}

\begin{figure}[h]
\centerline{\psfig{figure=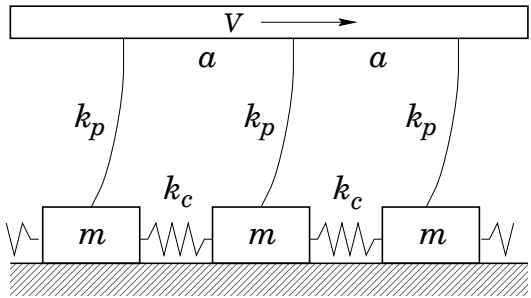,width=7cm}}
\caption{The Burridge-Knopoff model} \label{f1}
\end{figure}

The Burridge-Knopoff model consists of a one-dimensional array of
blocks of mass $m$ resting on a frictional surface
\cite{burridge67}. The blocks are connected together by springs with
spring constant $k_c$ and pulled by a loader plate moving with
constant speed $V$ via another set of springs having spring constants
$k_p$ (see Fig. \ref{f1}). 
Let us measure the displacement $X_i$ of
the $i$-th block relative to the point of attachment of the $i$-th
loader spring. In this case the total force $F_i$ acting on the $i$-th
block is given by
\begin{equation} \label{force0}
F_i = k_c \left( X_{i + 1} + X_{i - 1} - 2 X_i \right) - k_p X_i -
f_i,
\end{equation}
where $f_i$ is the force of friction. The dynamics of the system is
completely determined by the equation of motion $m \ddot{X}_i = F_i$,
provided that the friction law is specified. Note that the friction
force $f_i$ is the only nonlinearity in the equation of
motion. Clearly, the dynamics of the system will significantly depend
on the particular choice of the friction law. Recently, a lot of
results were presented on the dynamics of the Burridge-Knopoff model
in the case of velocity-weakening friction law
\cite{schmittbuhl93,espanol94,carlson89,carlson91,%
carlson94,langer91,myers93}. A characteristic feature of the
Burridge-Knopoff model with this form of friction is its highly
chaotic dynamics that occurs on all length scales down to the smallest
length scale $a$ and therefore, the absence of the proper continuum
limit \cite{carlson89}.

\begin{figure}
\centerline{\psfig{figure=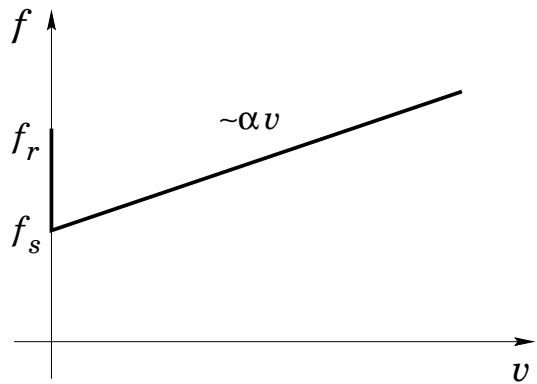,width=7cm}}
\caption{Friction force as a function of the velocity} \label{f2}
\end{figure}

In contrast to most previous studies, here we adopt the Coulomb
friction law, which is applicable to clean dry surfaces (see, for
example, \cite{bowden}). Namely, we will characterize the friction
force by the value of the static friction $f_r$ and the kinematic
friction $f_s < f_r$, where $f_r$ and $f_s$ are positive
constants. Thus, the block will remain at rest if $F_i < f_r$. When
$F_i$ reaches $f_r$ the block starts to move (slips) and the friction
force drops {\em instantaneously} from the value of $f_r$ to the value
of $f_s$. When the block comes to rest (sticks), static friction turns
on again. Also, we will supplement the friction force with the viscous
friction term $f_{\mathrm viscous}(v) = \alpha v$, where $v$ is the
velocity of the block and $\alpha$ is a constant (Fig. \ref{f2}). 
Note
that this kind of friction law was considered already in the original
paper of Burridge and Knopoff \cite{burridge67} and is observed in the
friction experiments and their analogs
\cite{heslot94,brechet94,kilgore93,lebyodkin95,field95}. Also note
that this model was recently studied numerically in Ref. \cite{xu94}
(although for the parameters different from those considered in this
paper).

We would like to emphasize that in contrast to velocity-weakening
friction law, for which the kinematic friction force is a continuous
function of the block velocity at $v = 0$, in our friction law the
force $f$ has a {\em discontinuity} at $v = 0$, reflecting the fact
that the coefficient of static friction is typically larger than that
of kinematic friction. As a result of this difference, the accumulated
stress {\em accelerates} the block at the onset of motion, releasing
potential energy. As we will see below, this provides the sustaining
force for the traveling waves studied in the following section. Also,
as we will show below, this ensures the existence of the proper
continuum limit as $a \rightarrow 0$.

Let us now formulate the equations of motion for the blocks in the
case of the friction law introduced above. Recall that the
displacements $X_i$ are measured relative to the loader plate, so they
are defined in the reference frame moving relative to the surface with
velocity $V$. Therefore, when the blocks are at rest, their equation
of motion becomes
\begin{equation} \label{rest0}
\dot{X}_i = - V.
\end{equation}
When the block slides, its equation of motion changes to
\begin{equation} \label{slide0}
m \ddot{X}_i = k_c ( X_{i + 1} + X_{i - 1} - 2 X_i) - k_p X_i - f_s -
\alpha ( \dot{X}_i + V).
\end{equation}
The transition to sliding (slip) occurs when the force $F_i$ reaches
the value of $f_r$:
\begin{equation} \label{slip0}
k_c ( X_{i + 1} + X_{i - 1} - 2 X_i) - k_p X_i = f_r.
\end{equation}
In other words, the equation of motion of a block changes from
Eq. (\ref{rest0}) to Eq. (\ref{slide0}) when the condition in
Eq. (\ref{slip0}) is satisfied.  Of course, the block comes back to
rest when $\dot{X}_i = -V$ during sliding.

Let us introduce the following dimensionless quantities
\begin{eqnarray} 
x' = {x \over a} \sqrt{ k_p \over k_c}, ~~~t' = t \sqrt{k_p \over m},
~~~u_i = - {k_p X_i + f_s \over f_r - f_s},
\label{dimless}
\end{eqnarray}
where $a$ is the distance between the attachment points of the loader
springs (Fig. \ref{f1}). The dimensionless displacement variable $u_i$
is chosen in such a way that the system is excitable only at $u_i > 0$
[see Eq. (\ref{force0}), in order for slip to occur, the force $F_i$
must exceed the value $f_s$ of sliding friction], while the blocks
will always slide for $u_i > 1$ [see Eq. (\ref{slip0})].

If $k_c \gg k_p$, one can naturally go to the continuum limit by
making a substitution $X_{i + 1} + X_{i - 1} - 2 X_i \rightarrow a^2
{\partial^2 X \over \partial x^2}$. This is a good approximation when
the length scale of variation of $X_i$ is of order 1 in the new
variables. Using the variables in Eq. (\ref{dimless}), we rewrite
Eqs. (\ref{rest0}) and (\ref{slide0}), respectively, as follows:
\begin{eqnarray}
u_t & = & v, \label{rest} \\ u_{tt} & = & u_{xx} - u - 2 \gamma (u_t -
v),
\label{slide} 
\end{eqnarray}
where we introduced dimensionless constants
\begin{equation}
\gamma = {\alpha \over 2 \sqrt{ k_p m}}, ~~~v = {V \sqrt{k_p m} \over
{f_r - f_s} },
\end{equation}
and dropped the primes for simplicity of notation. Note that in the
continuum limit in addition to tracking the motion of individual
blocks, one also has to follow the motion of the slip
points. Similarly, one should keep track of the positions of the
points at which the blocks come to rest. The latter are determined by
the condition in Eq. (\ref{rest}) during sliding.

Determining the position of the slip points in the continuum limit
turns out to be a rather complicated problem, since the position of
the slip will still depend on the local dynamics of the blocks. In the
discretized form, the slip condition [Eq. (\ref{slip0})] in the
variables of Eq. (\ref{dimless}) becomes
\begin{equation}
{u_{i+1} + u_{i-1} - 2 u_i \over (\Delta x)^2} = u_i - 1,
\label{slip}
\end{equation}
where $\Delta x = \sqrt{k_p/k_c}$. We will get back to this problem in
the following section where we will show that in the continuum limit
the dynamics of the slip becomes independent of $\Delta x$.

Equations (\ref{rest}) -- (\ref{slip}) are the constitutive equations
that will be studied in the rest of the paper. As can be seen from
Eq. (\ref{dimless}), we chose time and length scales so that the speed
of sound in the system is equal to 1. The time scale is determined by
the period of oscillations of an isolated block due to the loader
spring. Note that in the continuum limit the characteristic speed of a
block during sliding is much smaller than the speed of sound. This can
be seen from Eq. (\ref{dimless}) if one assumes $u_t \sim 1$, measures
$X_i$ in the units of Eq. (\ref{dimless}), and uses a natural
assumption that $f_r - f_s \ll a k_p$. The behavior of the system is
determined only by two parameters: the dimensionless dissipation
parameter $\gamma$ which measures the effect of viscous friction, and
the dimensionless rate of accumulation of stress $v$. In the following
we will consider $v$ to be small, what expresses the fact that the
time scale of accumulation of stress is much longer than that of the
motion of an individual block during sliding.

Finally, let us discuss the applicability of the Burridge-Knopoff
model in the context of real physical systems exhibiting stick-slip
motion. In a real system one should replace a one-dimensional array of
masses between the loader plate and the frictional surface by an
elastic medium of certain thickness $h$. A straightforward extension
of the Burridge-Knopoff model would therefore be a two-dimensional
array of masses connected by springs whose one edge is rigidly
attached to the loader plate and the other slides on the frictional
surface. Naturally, the thickness of such a medium should greatly
exceed the microscopic length scale $a$. The stick-slip motion in such
a medium is due to {\em surface waves} \cite{landau7}. The dispersion
of these waves is given by $\omega^2 = k^2 + (\pi / 2 h)^2 (1 + 2
n)^2$, where $n$ is an integer and the transverse speed of sound was
taken to be 1. The main difficulty here is that instead of a single
displacement variable $u$ one has to deal with a large number of modes
with different $n$. A simplification that is presented by the
Burridge-Knopoff model consists of lumping up these modes into a
single mode with an {\em effective} stiffness $k_p$
[Eq. (\ref{slide0})]. It is clear that the dominant modes will be
those with small $n$, so we must have $k_p \sim (a/h)^2 k_c \ll
k_c$. Thus, in a physically relevant situation one should consider the
continuum limit $\Delta x \rightarrow 0$ of the Burridge-Knopoff
model. In short, this is merely a simplification of the mathematical
handling of the elastic dynamics. A much more important assumption is
that the friction response to the motion of the medium is
instantaneous and has zero correlation length. It would be incorrect
to think of the ``size'' of the block $a$ as an atomic
distance. Rather, the value of the parameter $a$ should have a
magnitude of the correlation (memory) length of the cooperative
effects responsible for the surface friction (see, for example,
\cite{heslot94} and references therein). Then the response of the
system can be considered to be instantaneous if the characteristic
velocity of the blocks is greater than $a/\tau_{\mathrm stick}$, where
$\tau_{\mathrm stick}$ is the characteristic sticking time.

\section{traveling waves}

Let us now demonstrate that Eqs. (\ref{rest}) -- (\ref{slip}) admit
solutions in the form of traveling shock waves with constant speed $c$
in the limit of vanishingly small $v$. But before we do that, let us
see what kind of dynamics one would observe if there are no slip
events and the initial distribution of $u$ is taken to be a
sufficiently slowly varying distribution $u_0(x)$ (the latter ensures
that no slip events will occur at short times). In this case the
dynamics is trivial: we will have $u(x,t) = u_0(x) + v t$, as long as
no slip events occur [see Eq. (\ref{rest})]. From this one can see
that the characteristic time scale of accumulation of stress is
$v^{-1} \gg 1$, when $v$ is small. In other words, for $v \ll 1$ on
the time scale of order 1 one will not see any motion with the
distribution of $u$ fixed to $u_0(x) + C$, where $C$ is a constant.
In particular, if one starts with the uniform initial conditions, one
will have $u = u_+$, where $u_+$ is some constant less than 1. The
other possible situation when the dynamics of the system becomes
trivial is steady creep, when all the blocks move together with the
loader plate, so we simply have $u = 2 \gamma v \simeq 0$ for $v \ll
1$ [see Eq. (\ref{slide})].

Consider the system with $u = u_+$ and $v = 0$. As was noted in the
previous section, when $u_+ < 0$ slip events cannot occur, so if one
slightly moves a particular block, it will quickly move to a new
equilibrium position and no significant changes will occur. A
different situation is realized for $0 < u_+ < 1$ when the system
becomes excitable. Then, if a single block is slightly moved, it will
accelerate, releasing the potential energy accumulated in $u_+$. As a
result of this acceleration, the force acting on the adjacent blocks
will increase resulting in slipping of the adjacent block that was at
rest. This avalanche-like process will go on, leading to the formation
of the shock wave that releases the accumulated stress, so that $u$
changes to a new lower value of $u_-$. The situation here resembles a
great deal combustion of safety fuse discussed in the
Introduction. Thus, for $0 < u_+ < 1$ a small perturbation of $u$ may
lead to a significant change of the state of the system, switching it
from $u = u_+$ to $u = u_-$. As in other excitable systems
\cite{cross93,vasiliev,mikhailov,field,murray,ko:book}, it is natural
to expect that such a perturbation will transform at long times to a
shock wave traveling with constant velocity $c$. Numerical simulations
of Eqs. (\ref{rest}) -- (\ref{slip}) with $v = 0$ show that this
indeed happens in our model.

Since all the ``nonlinearity'' in the problem is contained in the slip
condition given by Eq. (\ref{slip}), our system is not unlike
piecewise-linear model used to study propagation of nerve impulses
\cite{rinzel73}. For this reason the solution in the form of the
traveling wave can be found exactly for $v = 0$ in the continuum
limit. For definiteness, we will consider the waves traveling from
left to right, so that $c > 0$. Because of the reflection symmetry,
for each solution traveling with speed $c$ there is also a solution
traveling with speed $-c$.

It is clear that the speed $c$ of the traveling wave should be
determined by the motion of the slip point in front of the wave. Let
the slip event occur for block $s$ at $t = t_s$. Then, for $i > s$ we
have $u_i = u_+$, so the slip condition from Eq. (\ref{slip}) becomes
\begin{equation} \label{slip1}
u_{s-1}(t_s) = u_+ - (1 - u_+) (\Delta x)^2.
\end{equation}
Since at the onset of the slip $u_s = u_+$ and $du_s/dt = 0$, at time
$t$ we will have $u_s = u_+ + {\cal O}\bigl( (t - t_s)^2 \bigr)$, with
$u_s < u_+$, since right after the slip the block is accelerated by an
excess force. This means that at some time $t_{s+1}$ the slip
condition from Eq. (\ref{slip}) will be satisfied for the $(s+1)$-th
block. According to Eq. (\ref{slip1}), this will happen when $ \Delta
t = t_{s+1} - t_s = {\cal O}(\Delta x)$. Therefore, on the time scale
of order 1 this will correspond to the motion of the slip point with
the speed $c = \Delta x / \Delta t = {\cal O}(1)$.

\begin{figure}
\centerline{\psfig{figure=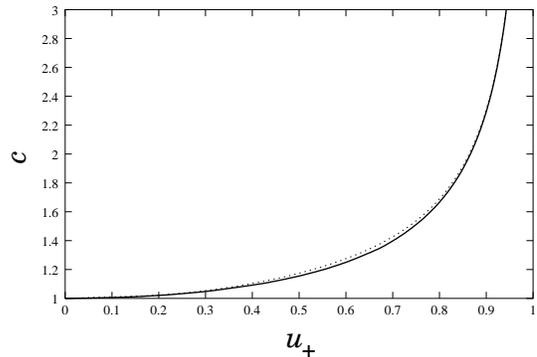,width=7cm}}
\caption{The dependence $c(u_+)$. Results of the numerical solution of
Eqs. (\ref{ini0}) -- (\ref{ini2}). Dashed line shows the result of
approximation by Eq. (\ref{c}). } \label{cf}
\end{figure}

To actually calculate the speed $c$ of the front, one needs to solve
the system of coupled equations of motion for the blocks behind the
$s$-th block. This problem in the limit $\Delta x \rightarrow 0$ is
considered in Appendix A. The analysis of this problem shows that the
speed $c$ of the wave is uniquely determined by the value of $u = u_+$
in front of the wave. Note that similar situation takes place in the
models of combustion (see, for example, \cite{fife}) and in
reaction-diffusion systems (see, for example,
\cite{cross93,vasiliev,mikhailov,field,murray,ko:book}). The
dependence $c(u_+)$ found numerically is shown in Fig. \ref{cf}. 
We would like to emphasize that upon the decrease of $\Delta x$ the
speed $c$ becomes independent of $\Delta x$, providing a well-defined
continuum limit for slip propagation. These results are also supported
by the direct numerical simulations of Eqs. (\ref{rest}) --
(\ref{slip}).

According to Fig. (\ref{cf}), traveling wave solution exists for all
$0 < u_+ < 1$. The speed $c$ is always greater than 1, that is, the
considered shock waves are supersonic. The latter is also observed in
other models of stick-slip motion \cite{langer91,myers93}. Note,
however, that the speed $c$ is the speed of {\em propagation} of the
slip and is not related with the actual speed of individual blocks in
the wave.

As can be seen from Fig. (\ref{cf}), the speed of the traveling wave
diverges as $u_+$ approaches 1. One can get an analytical handle on
the dependence $c(u_+)$ by expanding it in the inverse powers of $c$
(see Appendix A). As a result, we get the following interpolation
formula which implicitly determines $c$ as a function of $u_+$
\begin{equation} \label{c}
u_+ = \left( 1 + { 2 + c^2 + 4 c^4 \over 8 c^5 \sqrt{c^2 - 1}}
\right)^{-1}. 
\end{equation}
This equation gives the dependence $c(u_+)$ with a few per cent
accuracy. 

Let us introduce self-similar variable $z = x - c t$, where $c =
c(u_+)$ (Fig. \ref{cf}). Since the problem possesses translational
invariance, we can choose the position of the slip point to be at $z =
0$. Similarly, the stick point $z = -w$, with $w > 0$, will also
travel with speed $c$ behind the wave. Behind the slip the blocks
slide according to Eq. (\ref{slide}), so for the wave with speed $c$
we will have
\begin{equation} \label{slidet}
(c^2 - 1) u_{zz} - 2 c \gamma u_z + u = 0.
\end{equation}
The slip condition gives the following initial conditions for this
equation:
\begin{equation} \label{initc}
u(0) = u_+, ~~~u_z(0) = 0.
\end{equation}
The analysis of Eq. (\ref{slidet}) (see Appendix B) shows that the
structure of the solution changes qualitatively depending on whether
the value of $u_+$ is smaller or larger than the critical value $u_c >
0$, where the value of $u_c$ is implicitly determined by
\begin{equation} \label{gammac}
\gamma = {\sqrt{c^2(u_c) - 1} \over c(u_c)}.
\end{equation}

\begin{figure}[t]
\centerline{\psfig{figure=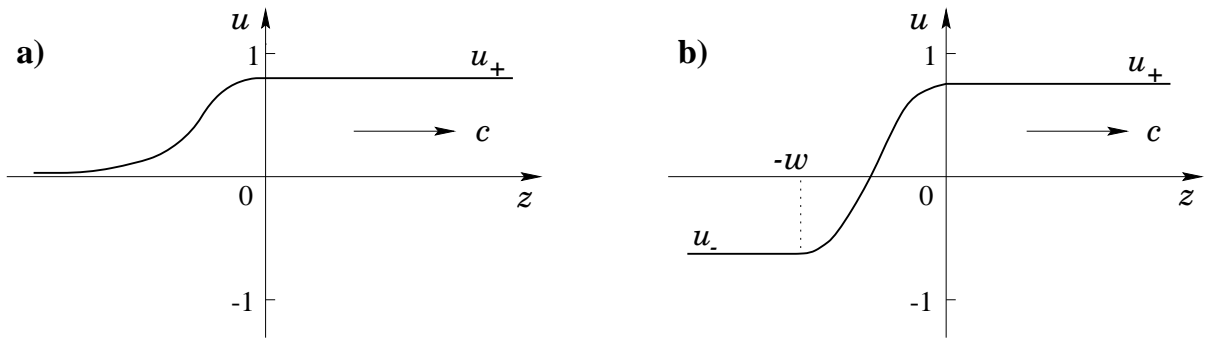,width=16cm,angle=-90}}
\caption{Two types of traveling wave solutions: (a) case $u_+ < u_c$;
(b) case $u_+ > u_c$. } \label{wave}
\end{figure}

If we have $u_+ < u_c$, the distribution of $u$ will asymptotically
approach $u_- = 0$ at $z = -\infty$ behind the wave
[Fig. \ref{wave}(a)]. 
\widetext
\narrowtext
In other words, the blocks behind the wave never
come to rest. Therefore, upon passing of the wave the system should
develop a steady creep for small but finite $v$. Note that this will
always happen when $\gamma > 1$ since in any case $u_+ < 1$, so for
these values of $\gamma$ the propagation of multiple shock waves
becomes impossible (see Sec. V).

A different situation takes place for $u_+ > u_c$. Then the solution
has the form shown in Fig. \ref{wave}(b), and the width $w$ of the
traveling wave becomes finite (see Appendix B)
\begin{equation} \label{w}
w = {\pi (c^2 - 1) \over \sqrt{ c^2 (1 - \gamma^2) - 1}}.
\end{equation}
Thus, in this case the wave propagates as a ``self-healing'' pulse
(compare with \cite{heaton90}). Note that for $\gamma > 0$ we have $c
> c(u_c) > 1$, so the value of $w$ is bounded from below and the width
of the wave is always greater or of order 1. This justifies the use of
the continuum limit for $k_c \gg k_p$ and $\gamma \sim
1$. Furthermore, when $u_+$ becomes close to 1, the propagation speed
$c$ and therefore the width of the wave grow, so in this case the
continuum limit is justified even for $k_c \sim k_p$. On the other
hand, the width $w$ goes to zero when $u_+$ becomes small when $\gamma
\ll 1$ [see Eq. (\ref{w})], so for sufficiently small $u_+$ and
$\gamma$ the continuum limit will become invalid for the description
of traveling waves.

In the case $u_+ > u_c$ the value of $u = u_-$ behind the wave will be
(see Appendix B)

\newpage

~\vspace{5.3cm}
\begin{equation} \label{um}
u_- = - u_+ \exp\left(- {\pi \gamma c \over \sqrt{c^2(1 -
\gamma^2) - 1}}\right).
\end{equation}
From this equation one can see that $u_- < 0$, so right behind the
wave the system is in the state in which no waves can be further
excited. Also, note that when $u_+$ approaches the value of
$u_c$, the value of $u_-$ rapidly approaches zero.

In the context of earthquakes a shock wave should be associated with
an individual earthquake. The total displacement $A = u_+ - u_-$ which
occurs upon passing of a wave and which determines the magnitude of
the earthquake will therefore be
\begin{equation} \label{A}
A = u_+ \left[ 1 + \exp\left(- {\pi \gamma c \over \sqrt{c^2(1 -
\gamma^2) - 1}}\right) \right].
\end{equation}

\section{free boundary problem}
 
So far we were studying traveling wave solutions in the limit $v = 0$
and $u = {\mathrm const}$ in front of the wave. These solutions are
clearly the {\em fast} motions in the system since they occur on the
time scales of order 1. On the other hand, although the accumulation
of stress occurs slowly for $v \ll 1$, it can lead to significant
changes in $u$ at long times of order $v^{-1} \gg 1$, so the latter
can be associated with the {\em slow} motions in the
system. Therefore, since the fast motions result in the large-scale
changes in $u$, the slip-stick events can be considered as {\em
singular perturbations} to the slow motions.

Similar situation is realized in other excitable media where the role
of singular perturbation is played by the sharp fronts
\cite{cross93,vasiliev,mikhailov,field,murray,ko:book}. A powerful tool in
describing the dynamics of such systems is reduction to free boundary
problem (see, for example, \cite{fife}). In this approach one obtains
the relationship between the speed of the sharp front and slow
variables. Using this idea, let us formulate the dynamics of our
system as a free boundary problem. 

Let us assume that the distribution of $u$ varies on the length scale
much greater than 1. Then, introducing an adiabatic approximation, we
can assume that the speed of the traveling wave is given by the
dependence $c(u_+)$ (Fig. \ref{cf}) in which now, instead of $u_+$ one
should use an instantaneous value of $u$ right in front of the
wave. The latter will play the role of the slow variable. Thus, the
motion of the shocks can be described by the variables $x_i(t)$ which
give the positions of the $i$-th shock, and the variables $s_i = \pm
1$ which give the direction of their motion.

On large length scales the shock will contribute a discontinuity to
the distribution of $u$ in the limit $v \rightarrow 0$. It can be
included in Eq. (\ref{rest}) describing slow motions in the form of a
$\delta$-function. Therefore, in the presence of the shocks
Eq. (\ref{rest}) can be rewritten as
\begin{equation} \label{slow}
u_t = v - \sum_i  c_i A(u) \delta (x - x_i + 0 s_i),
\end{equation}
where $A$ is the amplitude of the shock from Eq. (\ref{A}) evaluated
at $u_+ = u(x_i(t),t)$, $c_i$ is the absolute value of the speed of
the $i$-th shock, and the last term in $\delta$-function represents
the fact that the value of $u$ is evaluated right in front of the
wave. Equation (\ref{slow}) simply says that at the moment $t$ the
value of $u$ jumps from $u(x_i(t), t)$ at $x = x_i(t)$ to the new
value $u_-(u(x_i(t), t))$ given by Eq. (\ref{um}).

Having now defined the evolution of the slow variable $u$, we can
write down the equation of motion of the shocks in terms of it
\begin{equation} \label{fast}
\dot{x}_i = s_i c_i,
\end{equation}
where $c_i$ is the function of $u_+ = u\bigl(x_i(t), t\bigr)$
(Fig. \ref{cf}).

Equations (\ref{slow}) and (\ref{fast}) are the basic equations of the
free boundary problem describing the dynamics of the system for $v \ll
1$ and sufficiently slowly varying initial conditions, assuming that
no creep occurs in the system during its evolution. These equations
have to be supplemented with the way of dealing with creation and
annihilation of shocks. From the physical considerations it is clear
that a pair of shocks moving toward each other will
annihilate. Similarly, when the value of $u$ reaches 1 [recall that
our distributions of $u$ vary sufficiently slowly, so one can ignore
the variation of $u$ in Eq. (\ref{slip})], a slip event is initiated
at the point where $u = 1$, creating a pair of counter-propagating
shocks. These features should be incorporated into the free boundary
problem and will be discussed in the following section. Notice that
according to Eq. (\ref{slip}) the distribution of $u$ for blocks at
rest must be a continuously differentiable function in the continuum
limit, so the speeds $c_i$ of the shocks given by Eq. (\ref{fast})
will also be continuously differentiable functions of time.

In writing Eq. (\ref{slow}) we assumed that we always have
$u\bigl(x_i(t), t\bigr) > u_c$, so that the shock waves are always the
waves of switching between the blocks with $u = u_+$ at rest in front
of the wave to $u = u_-$ at rest behind the wave. In other words, we
do not consider the possibility of the onset of creep behind the
wave. The analysis of creep motion is beyond the scope of the present
paper. The condition of the absence of creep should in fact be
satisfied for a wide class of initial conditions. In particular, it is
easy to see that a wave with $u_+ > u_c$ will never be able to reach
the points where $u < u_c$ if we have
\begin{equation} \label{nocreep}
|u_x| < {v \over c(u_c)},
\end{equation}
provided that the regions with $u < u_c$ are initially at rest.  The
condition in Eq. (\ref{nocreep}) simply means that the wave with speed
$c$ will never catch up with the point at which $u = u_c$ when $u$
grows according to Eq. (\ref{rest}).

\section{spatiotemporal patterns}

\begin{figure}
\centerline{\psfig{figure=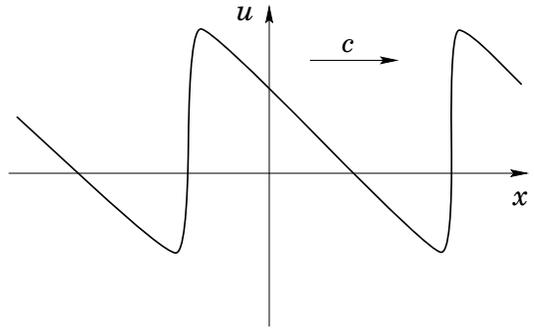,width=7cm}}
\caption{Periodic wave train. } \label{period}
\end{figure}

Let us now apply the procedure developed in the preceding section to
spatiotemporal patterns forming in the system under consideration. As
in any excitable system
\cite{cross93,vasiliev,mikhailov,field,murray,ko:book}, the basic
pattern of this kind is a periodic array of shocks (wave train)
traveling with constant velocity. From the results of the preceding
section one can see that a periodic wave train should consist of sharp
fronts in which the value of $u$ jumps from $u_+$ to $u_-$ followed by
refractory regions where $u$ slowly recovers back to $u_+$, so the
resulting pattern is saw-tooth (Fig. \ref{period}). 
In the refractory
region $u$ obeys Eq. (\ref{rest}), which in the frame moving together
with the wave with speed $c$ becomes $u_z = -v/c$, where $z = x - ct$
is the self-similar variable. Therefore, the solution for $u$ in the
refractory region which properly matches with the back of one shock at
$z = 0$ and the front of another at $z = -L$, where $L$ is the period
of the wave train, has the form $u = L^{-1}(z + L) u_- - L^{-1}z u_+$,
where
\begin{equation} \label{L}
L = {c A \over v},
\end{equation}
$c = c(u_+)$ (Fig. \ref{cf}), and $A$ is given by
Eq. (\ref{A}). Equation (\ref{L}) should in fact be obvious from the
purely geometric considerations. Thus, there exists a family of
nonlinear periodic traveling wave solutions whose speed and period
depend strongly on amplitude. Similarly, the period $T$ of this
solution in time depends on the amplitude simply as $T = A/v$, so the
frequency of the shocks at a particular point as the wave train passes
is $\omega \sim A^{-1}$. The latter in fact represents a simple form
of the Gutenberg-Richter scaling law with scaling exponent $b = 1$
\cite{gutenberg65}. In other words, this exponent would be observed in
the system under consideration if its dynamics were dominated by
periodic wave trains. It is interesting to note that the value of $b$
actually observed from the earthquakes generally lies in the range
$0.8 < b < 1.2$ \cite{evernden70}.

The above arguments are justified as long as $u_+ > u_c$, so that no
creep develops behind the traveling front. Therefore, according to
Eqs. (\ref{A}), and (\ref{L}), traveling wave trains exist only when
$L > L_{\mathrm min}$, where
\begin{equation} \label{Lmin}
L_{\mathrm min} = v^{-1} c(u_c) u_c.
\end{equation}

\begin{figure}
\centerline{\psfig{figure=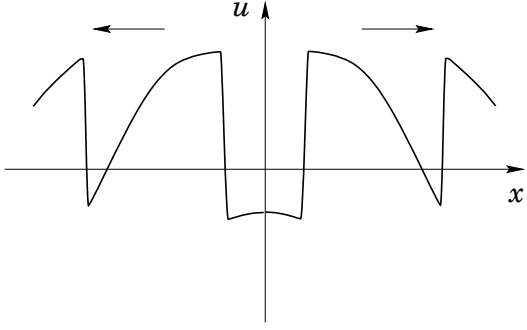,width=7cm}}
\caption{A source of counter-propagating waves.}
\label{guide}
\end{figure}

Let us now consider another kind of spatiotemporal patterns that is
typical of excitable systems --- the source of counter-propagating
waves \cite{cross93,vasiliev,mikhailov,field,murray,ko:book}. The
existence of such a solution is associated with the fact that if there
is an inhomogeneous distribution of $u$ at rest such that the maximum
of $u$ is located at $x = x_0$, then at some point in time the value
of $u(x_0)$ may reach 1, so that the blocks will become unstable
creating a pair of counter-propagating shocks
(Fig. \ref{guide}). 
After these shocks moved away from $x_0$, the
distribution of $u$ can once again have a maximum at $x = x_0$, so
after some time the value of $u$ will increase until it reaches 1 and
the cycle repeats. 

These solutions are indeed realized in our system. Close to $x = x_0$
we can approximate the distribution of $u$ as
\begin{equation} \label{a}
u(x, t) \simeq 1 - a (x - x_0)^2 + vt,
\end{equation}
[see Eq. (\ref{rest})], where we also assumed that the slip occurs at
$t = 0$. This time-dependent distribution of $u$ gives the values of
$u$ in front of the shocks at $t > 0$, after the slip occurred, thus
determining the velocity of the shocks. The analysis of the free
boundary problem in this case (Appendix C) then shows that right after
the slip the positions of the shocks will be given by
\begin{equation} \label{b}
x_i \simeq x_0 \pm \sqrt{bt}, ~~~~~b = {v + \sqrt{v^2 + 8 a} \over 2
a},
\end{equation}
where the plus and the minus correspond to the shocks propagating to
the right and to the left, respectively. Notice that Eq. (\ref{b}) in
general shows how to treat the singularity in the free boundary
problem (Sec. IV) associated with the creation of a a pair of
counter-propagating shocks. 

\begin{figure}
\centerline{\psfig{figure=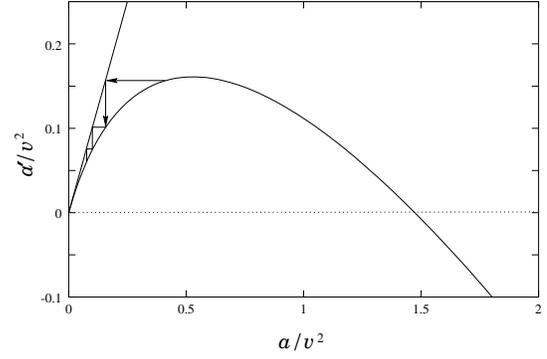,width=7cm}}
\caption{Dependence $a'(a)$ from Eq. (\ref{ap}) at $\gamma = 0.3$. }
\label{aa}
\end{figure}

The solution for $x_i(t)$ in Eq. (\ref{b}) then allows to calculate
the distribution $u'$ after the shocks have passed
\begin{equation} \label{up}
u'(x, t) \simeq u_-(1) - a' (x - x_0)^2 + vt,
\end{equation}
where $a'$ is given by (see Appendix C)
\begin{equation} \label{ap}
a' = -a \kappa + {2 a v (1 + \kappa) \over v + \sqrt{v^2 + 8 a}},
\end{equation}
$u_-(1) = - \exp(-\pi \gamma / \sqrt{ 1 - \gamma^2})$ [see
Eq. (\ref{um})], and
\begin{equation} \label{kappa}
\kappa = \left( 1 + {\pi \gamma \over (1 - \gamma^2)^{3/2}} \right)
\exp \left(-{\pi \gamma \over \sqrt{1 - \gamma^2}} \right).
\end{equation}
From Eq. (\ref{kappa}) one can see that the value of $\kappa$ lies in
the range $0 < \kappa < 1$.  The plot of $a'$ as a function of $a$ at
a particular value of $\gamma$ is presented in Fig. \ref{aa}.

From Eq. (\ref{ap}) one can see that there is a maximum of $u'$ at $x
= 0$ ($a' > 0$) when the shocks have passed, if $a < a_{\mathrm max}$,
where
\begin{equation} \label{amax}
a_{\mathrm max} = {v^2 (1 + \kappa) \over 2 \kappa^2}.
\end{equation}
If this condition is satisfied, after time $T = v^{-1} \left( 1 + \exp
\bigl(-\pi \gamma / \sqrt{1 - \gamma^2})\bigr) \right)$ the
distribution of $u$ will look exactly like the one in Eq. (\ref{a})
which we started with, but with a different value of $a$. Therefore,
Eq. (\ref{ap}) defines an iterative map for the value of $a$
corresponding to the solutions for $u$ at times $n T$, where $n$ is
the number of iterations. Physically, a source with period $T$ will
form at $x = 0$ creating traveling waves propagating away from
it. Note, however, that even though the period of the source is a
constant that does not depend on the initial conditions, the solution
represents an aperiodic process since after each cycle the value of
$a$ changes. The latter is represented in Fig. \ref{aa}. Observe that
we always have $a' < a$. From Fig. \ref{aa} one can also see that
after many periods the value of $a$ will tend to zero. The analysis of
Eq. (\ref{ap}) (see also Fig. \ref{aa}) shows that zero is the only
fixed point of the map $a \rightarrow a'$ for any $\gamma$. For small
enough values of $a$ Eq. (\ref{ap}) can be approximated as $a' \simeq
a[1 - 2 a (1 + \kappa)/v^2]$.  It is then easy to show that after $n
\gg 1$ iterations we will have $a_n \simeq v^2 / [2 n (1 + \kappa)]$,
where $a_n$ is the result of $n$ iterations of the map. Note that
because of the fact that $a$ becomes smaller and smaller with each
cycle, the distribution of $u$ becomes flatter and flatter [see
Eq. (\ref{a})], what means that after a long time the regions in the
neighborhood of the source will tend to synchronize. The
characteristic size $R$ of a region in which the synchronization takes
place can be estimated as $R \sim a_n^{-1/2}$, so we will have $R \sim
t^{1/2}$. Also note that for the same reason if there are two sources
with unequal values of $a < a_{\mathrm max}$ some distance $R$ apart,
after time $t \sim R^2$ the one with the smaller value of $a$ will
overwhelm the one with the larger value of $a$. It is then natural to
expect that if the initial distribution of $u$ has all maxima with $a
< a_{\mathrm max}$, the system will eventually synchronize into
uniform oscillations in which all the blocks slide at once.

Different situation is realized when $a > a_{\mathrm max}$. In this
case the distribution of $u$ forming after the shocks moved away from
the origin will be a {\em minimum} rather than a maximum of $u$, so
the next creation of a pair of shocks will not occur at $x = x_0$, but
rather at different points in space. Therefore, it is natural to
expect that for sufficiently random initial conditions one would see
shocks created and annihilated at random points in space, creating a
kind of spatiotemporal chaos.

\section{conclusion}

Let us now summarize the results of our analysis and discuss the
relationship of our results with those obtained for other models of
stick-slip motion. In the present paper we analytically investigated
the Burridge-Knopoff slider-block model with Coulomb friction law,
which is different from the one used in most previous studies
\cite{schmittbuhl93,espanol94,carlson89,carlson91,carlson94,%
langer91,myers93}. This difference is expressed in the fact that our
friction is a discontinuous function when the velocity of the block
becomes zero. As a result, the system is capable of propagating
ultrasonic traveling solitary waves in the limit of zero loader plate
velocity for any value of accumulated stress above the excitability
threshold $u = 0$. The existence of such solutions is a characteristic
feature of excitable systems
\cite{cross93,vasiliev,mikhailov,field,murray,ko:book}.

In the continuum limit the solution in the form of the traveling shock
wave is independent of the short-scale behavior of the system, except
for the precise form of the dependence of the wave's velocity on the
amount of the accumulated stress in front of the wave. The latter in
fact only weakly depends on the dynamics of the model at short length
scales if the value of $u$ is sufficiently close to the slipping
threshold $u = 1$. Note that one might want to calculate the
dependence of the wave velocity on the amount of the accumulated
stress $u_+$ by formally combining Eq. (\ref{slidet}) at $z = 0$ with
the continuum version of Eq. (\ref{slip}) at $z = 0$. This would give
an incorrect result $c = 1 / \sqrt{1 - u_+}$. This means that although
the continuum approximation may be valid for finding the wave profile
behind the slip point, one still needs to solve the discrete problem
at the tip of the wave in order to find its velocity.

In the limit of vanishing loader plate velocity the only parameter in
the model is the dimensionless coefficient of viscous friction
$\gamma$. All our analysis was performed for arbitrary values of
$\gamma$, so it is also applicable to the case $\gamma = 0$, i.e.,
when the viscous friction is absent. This case, however, has one
special feature. The thing is that for $\gamma = 0$ we have $u_c = 0$
[see Eq. (\ref{gammac})], so as $u$ approaches zero the speed of the
wave approaches 1 and the width $w$ goes to zero [see
Eq. (\ref{w})]. Therefore, for a fixed value of $\Delta x \ll 1$ the
continuum limit will no longer be justified for the description of the
waves when $u$ becomes sufficiently close to zero. Also, when $\gamma
= 0$ the amplitude $A = 2 u_+$ of these waves [see Eq. (\ref{A})] will
be able to become arbitrarily small.

According to the results of our analysis, the solution in the form of
the traveling shock wave, as well as a periodic wave train of a given
period, is unique in the limit of vanishingly small loader plate
velocity, and its speed is uniquely determined by the value of $u$ in
the front of the wave. This is in contrast with the results of Langer
and Tang \cite{langer91} and Myers and Langer \cite{myers93} for the
models with velocity-weakening friction where they find a continuous
family of such solutions and therefore a selection problem.

Note that the systems with velocity-weakening friction transform to
the model considered by us in the limit of small $V_f$, where $V_f$ is
the characteristic velocity of variation of the friction
force. Indeed, if the characteristic speed of a block during the slip
exceeds $V_f$, the friction force will drop very rapidly, acting in
the same way as in our model. To obtain the criterion of smallness of
$V_f$, we note that the effect of the velocity dependence of the
friction is most pronounced in the slip region, where, according to
Eq. (\ref{slip1}), $u_t \sim \Delta x = \sqrt{k_p / k_c}$, where we
assumed $c \sim 1$, so $u$ is not near the threshold value $u = 1$. In
the original variables $d X_i/dt \sim (f_r - f_s) / \sqrt{k_c m}$, so
the condition $V_f \ll d X_i/ d t$ gives
\begin{equation} \label{crit}
V_f \ll {f_r - f_s \over \sqrt{k_c m}}.
\end{equation}
This formula should in fact be obvious from the physical
considerations. Similarly, one should recover our results for the
creep-slip models considered in \cite{cartwright97} if the
characteristic velocity of variation of the friction force is small
enough.

In contrast, if the condition opposite to the one in Eq. (\ref{crit})
holds, no traveling shock waves can be realized in the continuum
limit. This can be seen from the following argument. In a traveling
wave the profile of $u$ should be described by Eq. (\ref{slidet}) in
which, in the case of the velocity-weakening friction one should drop
the term $-2 c \gamma u_z$ and replace $u$ by $u-1$ close to the slip
point. There we must have $u_{zz} < 0$, so, according to the modified
Eq. (\ref{slidet}), we will have $c < 1$ in the traveling wave. On the
other hand, the propagation of subsonic waves in the system is
impossible since from the dispersion relation $\omega^2 = 1 + k^2$ for
the waves in the absence of friction we get that their phase velocity
$c = \omega / k = \sqrt{1 + k^{-2}} > 1$ for all $k$. On the other
hand, numerical simulations show that in the discrete models with
velocity-weakening friction traveling shock waves are indeed
observed. However, for $\Delta x \ll 1$ these waves move with speeds
very close to the speed of sound and their width is just a few lattice
spacings (unless one is close to the threshold $u = 1$, see also
\cite{langer91,myers93}). One should therefore expect that in the
continuum limit only discontinuous shocks traveling with the speed of
sound are feasible, in contrast to the situation studied in the
present paper. This is a crucial distinction of the Burridge-Knopoff
model with velocity-weakening friction and the one with Coulomb
friction law (Fig. \ref{f2}) studied by us.

As was already discussed above, the propagation speed of the wave
depends on the short-scale dynamics of individual blocks. Therefore,
in order to be able to perform numerical simulations of the models
under consideration in the continuum limit, one has to use very small
discretization of time. This should make direct simulations of our
model rather difficult. On the other hand, we showed that if the
initial conditions vary sufficiently slowly, the dynamics of the
system reduces to the motion of individual shock waves (see
Sec. IV). Therefore, by reformulating the dynamics of the problem in
terms of the free boundary problem one can dramatically increase the
speed of the simulations, thus being able to observe the dynamics for
much longer times. This should be important in the statistical
analysis of our system, especially in the context of earthquakes (see
also \cite{carlson89,carlson91,carlson94}).

Let us see how the results obtained from the studies of our model
should translate to real stick-slip problems. Consider, for example, a
Bristol board of thickness $h = 2$ mm lying on the surface made of the
same material under external pressure $p$ and pulled at the top
\cite{heslot94}. The coefficients $\mu_r$ and $\mu_s$ of the static
and the sliding friction for these surfaces were found to be $\mu_r =
0.37$ and $\mu_s = 0.31$, respectively \cite{heslot94}. The memory
length which we should use for the distance $a$ between the blocks was
found to be $a = 1~\mu$m. A single block in the Burridge-Knopoff model
should be associated with a piece of the board of dimensions $a \times
a \times h$, so the mass $m$ of the block should be $m = 2.4 \times
10^{-12}$ kg, where we used the density $\rho = 1.2~{\mathrm
g/cm}^3$. From the speed of sound $c = a \sqrt{k_c/m}$
[Eq. (\ref{slide0})] we can calculate the values of $k_c$ and $k_p =
k_c (\pi a / 2 h)^2$ (see the end of Sec. II). Using $c = 4000$ m/s,
we find $k_c = 3.8 \times 10^{7}$ N/m, and $k_p = 2.3 \times 10^1$
N/m. The friction force jump per single block is $f_r - f_s = p
a^2(\mu_r - \mu_s)$. One can see that the condition of
Eq. (\ref{crit}) with $V_f = 10^{-5}$ m/s (see \cite{heslot94}) is
satisfied if $p = 2 \times 10^6 {\mathrm ~Pa} = 20$ atm, what gives
$f_r - f_s = 1.2 \times 10^{-7}$ N. For this value of $p$ the
characteristic sliding velocity will be $\dot{X} \sim (f_r - f_s)/
\sqrt{k_p m} = 1.6$ cm/s. The displacement $\Delta X$ in a single wave
can also be estimated as $\Delta X = 2 (f_r - f_s)/k_p = 1 \times
10^{-8}$ m. The latter quantity turns out to be quite small. Note that
both $\Delta X$ and $\dot{X}$ increase as $p$ increases. Also, smaller
pressures $p$ would be needed for smaller values of $h$. From
\cite{heslot94} one can calculate $\alpha = 6.7 \times 10^{-6}$ kg/s
for this value of $p$, leading to $\gamma = 0.45$.

In the present paper we studied perfectly homogeneous system without
any external noise. It is clear that in real systems some degree of
noise should be present. It is interesting to know what effect the
introduction of randomness will have on the behavior of the traveling
waves. Here one can distinguish two different situations. If one adds
some small randomness in the parameters of individual blocks, such as
the friction coefficients, spring constants and so forth, one would
expect the renormalization of the speed of the traveling waves. The
other kind of noise would result in sudden transitions of the blocks
at rest to sliding. This kind of noise can be readily incorporated
into our free boundary problem by introducing generation of pairs of
counter-propagating waves at random points in space. Here the open
question is how the frequency of this generation should depend on the
distance to the slipping threshold. It is also interesting how the
presence of such a noise would affect the wave statistics.

\appendix

\section{}

Since the variation of $u_i$ leading to the slip event is of order
$\Delta x^2$ and is small for small $\Delta x$, for the blocks near
the slip point one can neglect both the variation of $u_i$ and the
term $-2 \gamma du_i / dt$ in Eq. (\ref{slide}), so in the limit
$\Delta x \rightarrow 0$ the right-hand side of this equation simply
becomes $u_+ - 1 = {\mathrm const}$. For convenience, let us introduce
the new variables $\tau$ and $v_i$, such that
\begin{equation} \label{nv}
t = t_s + \tau \Delta x, ~~~u_i = u_+ - u_+ (\Delta x)^2 \left( v_i +
{\tau^2 \over 2} \right).
\end{equation}
Since at $\tau = 0$ the $s$-th block just slipped, we
must have [see Eq. (\ref{slip1})]
\begin{equation} \label{slip2}
v_{s-1} (0) = {1 - u_+ \over u_+},
\end{equation}
with the initial conditions
\begin{equation} \label{ini0}
v_s(0) = 0, ~~~~v'_s(0) = 0, 
\end{equation}
where the prime denotes the derivative with respect to $\tau$.

In the limit $\Delta x \rightarrow 0$ Eq. (\ref{slide}) can be written
as
\begin{eqnarray}
{d^2 v_i \over d \tau^2} = v_{i+1} + v_{i-1} - 2 v_i, && ~~~~i < s,
\label{elast1} \\ {d^2 v_i \over d \tau^2} = v_{i-1} - 2 v_i - {\tau^2
\over 2}, && ~~~~i = s. \label{elast2} 
\end{eqnarray}
These equations essentially describe a linear array of unit masses
connected by springs with spring constant 1, with the $s$-th mass
attached to a rigid wall and acted upon by a time-dependent force
$-\tau^2/2$. 

If the wave moves with speed $c$, after time $\tau = 1/c$, which
corresponds to $\Delta t = \Delta x/c$ [see Eq. (\ref{nv})], it should
move the distance $\Delta x$ to the right. This means that at time
$t_{s+1} = t_s + \Delta t$ we must have $u_{i-1}(t_s) = u_i(t_s +
\Delta t)$, $u'_{i-1}(t_s) = u'_i(t_s + \Delta t)$, or in terms of the
new variables
\begin{eqnarray}
v_{i-1} (0) & = & v_i (c^{-1}) + { 1 \over 2 c^2}, \label{ini1} \\
v'_{i-1} (0) & = & v'_i (c^{-1} ) + {1 \over c}. \label{ini2}
\end{eqnarray}
Note that these conditions introduce a feedback into
Eqs. (\ref{elast1}) and (\ref{elast2}). Also, the solution of
Eqs. (\ref{elast1}) should be matched with the continuum profile
behind the tip, which satisfies Eq. (\ref{slidet}). We should
therefore have
\begin{equation}
v_{s-k} \cong {k^2 \over 2 (c^2-1)}, ~~~k \gg 1.
\end{equation}

Equations (\ref{ini0}) -- (\ref{ini2}) can be solved numerically by an
iterative procedure. Then, calculating the value of $v_s(c^{-1})$,
using Eq. (\ref{ini1}) as a function of $c$, one can relate it to
$u_+$ through Eq. (\ref{slip2}). The results of this numerical
solution for $c > 1$ are presented in Fig. (\ref{cf}). These results
also show that close to $u_+ = 0$ we have
\begin{equation} \label{c1}
c \simeq 1 + {1 \over 2} u_+^2.
\end{equation}

The problem in Eqs. (\ref{ini0}) can be treated analytically by
expanding its solution in the powers of $c^{-1}$. Indeed, if $c$ is
large, for $\tau < c^{-1} \ll 1$ the right-hand sides in
Eqs. (\ref{elast1}) and (\ref{elast2}) can be neglected, so the
solution of Eqs. (\ref{ini0}) -- (\ref{ini2}) for $k \geq 0$ in the
leading order in $c^{-1}$ will be
\begin{eqnarray}
v_{s-k}^{(0)} = {k^2 \over 2 c^2} + {k \tau \over c}.
\end{eqnarray}
Physically, this corresponds to the situation in which the springs
between the blocks are absent.

To calculate the first order correction $v_s^{(1)}$, we use
$v_{s-1}^{(0)}$ in Eq. (\ref{elast2}) to obtain
\begin{equation}
v_{s}^{(1)} = {\tau^2 \over 4 c^2} + {\tau^3 \over 6 c} - {\tau^4
\over 24}.
\end{equation}
This solution is then used to fix through Eqs. (\ref{ini1}) and
(\ref{ini2}) the initial condition in Eq. (\ref{elast1}) with $i =
s-1$ to the next order in $c^{-1}$. Using the solutions $v_s^{(0)}$,
$v_{s-1}^{(0)}$, and $v_{s-2}^{(0)}$ in Eq. (\ref{elast1}), one can
then calculate $v_{s-1}^{(1)}$:
\begin{equation}
v_{s-1}^{(1)} = {3 \over 8 c^4} + {5 \tau \over 6 c^3} + {\tau^2 \over
2 c^2}. 
\end{equation}
The knowledge of $v_{s-1}^{(1)}$ and $v_s^{(1)}$ then allows to
calculate the next order correction $v_s^{(2)}$ by substituting them
into Eq. (\ref{elast2})
\begin{equation}
v_s^{(2)} = {3 \tau^2 \over 16 c^4} + {5 \tau^3 \over 36 c^3} -
{\tau^5 \over 60 c} + {\tau^6 \over 360}.
\end{equation}
Substituting the obtained solution for $v_s$ into Eq. (\ref{slip2})
and using Eq. (\ref{ini1}) with $i = s-1$, we get
\begin{equation} \label{ca}
{1 \over 2 c^2} + {3 \over 8 c^4} + {5 \over 16 c^6} = {1 - u_+ \over
u_+}.
\end{equation}
Note that this procedure can be continued to an arbitrary order in
$c^{-1}$, although the calculation then becomes rather cumbersome. Let
us only quote the result that the correction to the left-hand side of
Eq. (\ref{ca}) due to the next order terms will be $0.2719/c^8$.

Finally, combining Eq. (\ref{ca}) with Eq. (\ref{c1}), we arrive at
the interpolation formula given by Eq. (\ref{c}).

\section{}

Here we solve Eq. (\ref{slidet}) with the initial conditions given by
Eq. (\ref{initc}) for arbitrary $\gamma$. Let us introduce the new
variable $\xi = z / \sqrt{c^2 - 1}$ and a constant $\tilde\gamma = {c
\gamma / \sqrt{c^2 - 1}}$, where $c$ is a function of $u_+$
(Fig. \ref{cf}). Then, Eq. (\ref{slidet}) can be rewritten as
\begin{equation} \label{slidexi}
u_{\xi\xi} - 2 \tilde\gamma u_\xi + u = 0,
\end{equation}
with the same initial conditions written in terms of $\xi$:
\begin{equation} \label{initxi}
u(0) = u_+, ~~~~~u_\xi(0) = 0.
\end{equation}

Two situations are possible here. If we have $\tilde\gamma > 1$, both
roots of the characteristic equation of Eq. (\ref{slidexi}) are real,
so the solution has the form 
\begin{equation} \label{real}
u = a e^{\xi (\tilde\gamma + \sqrt{\tilde\gamma^2 - 1})} + b e^{\xi
(\tilde\gamma - \sqrt{\tilde\gamma^2 - 1})}.
\end{equation}
Using the initial conditions from Eq. (\ref{initxi}), we obtain for
the coefficients $a$ and $b$
\begin{equation}
a = {u_+ \over 2} \left( 1 - {\tilde\gamma \over \sqrt{\tilde\gamma^2
- 1}} \right), ~~~~~b = {u_+ \over 2} \left( 1 + {\tilde\gamma \over
\sqrt{\tilde\gamma^2 - 1}} \right).
\end{equation}
From the solution we see that behind the wave the distribution of $u$
asymptotically approaches zero. This means that we have $u_- = 0$, but
the system comes to rest only at $z = -\infty$, so the width $w$ of
the wave is infinite. The form of the solution is shown in
Fig. \ref{wave}(a). 

In contrast, when $\tilde\gamma < 1$, the roots of the characteristic
equation of Eq. (\ref{slidexi}) become complex conjugate, so the
solution has the form
\begin{equation} \label{complex}
u = a e^{\tilde\gamma \xi} \cos (\xi \sqrt{1 - \tilde\gamma^2}) + b
e^{\tilde\gamma \xi} \sin (\xi \sqrt{1 - \tilde\gamma^2}).
\end{equation}
Substituting the initial conditions from Eq. (\ref{initxi}), we obtain
\begin{equation} \label{ab}
a = u_+, ~~~~~b = - {u_+ \tilde\gamma \over \sqrt{1 -
\tilde\gamma^2}}. 
\end{equation}
In constructing the full solution one should also take into account
that the block will stick back when it comes to rest, what will happen
when $u_\xi = 0$ for vanishing $v$. According to Eqs. (\ref{complex})
and (\ref{ab}), this will occur at $z = z_s = - w$, where
\begin{equation}
w = {\pi \over \sqrt{1 - \tilde\gamma^2}}.
\end{equation}
Behind the stick point the value of $u$ will remain constant equal to
$u_-$, with
\begin{equation}
u_- = - u_+ \exp \left( - {\pi \tilde\gamma \over \sqrt{1 -
\tilde\gamma^2}} \right).
\end{equation}
Going back to $\gamma$, we can rewrite the equations above as
Eqs. (\ref{w}) and (\ref{um}), respectively. The form of the solution
in this case is shown in Fig. \ref{wave}(b).

The solutions obtained above are the exact traveling wave solutions of
Eqs. (\ref{rest}) -- (\ref{slip}) in the case $v = 0$. As can be seen
from the construction, traveling wave solution is unique for any given
value of $u_+$.

\section{}

When a pair of counter-propagating shocks is created, their speed will
be large, since in the neighborhood of $x_0$ the value of $u$ will be
close to 1. This allows us to use Eq. (\ref{ca}) and write
\begin{equation} \label{cc}
c_i \simeq {1 \over \sqrt{2[1 - u(x_i, t)]}}
\end{equation}
for $x_i$ close to $x_0$. Note that Eq. (\ref{cc}) expresses the fact
that for $u$ close to 1 the coupling of the blocks through the
longitudinal springs becomes inessential (see appendix A). 

Without any loss of generality, we can put $x_0 = 0$. According to
Eqs. (\ref{fast}) and (\ref{cc}), we have approximately
\begin{equation} \label{aaa}
\dot{x}_i = \pm {1 \over \sqrt{2 (a x_i^2 - v t} ) }
\end{equation}
for not too large $|x_i|$ and $t$, with the initial condition $x_i(0)
= 0$. The solution $x_i(t)$ that satisfies this equation and the
initial condition has the form $x_i = \pm \sqrt{b t}$, where the
constant $b$ has to be determined. Substituting this expression for
$x_i(t)$ into Eq. (\ref{aaa}), we obtain the following quadratic
equation for $b$:
\begin{equation}
a b^2 - v b - 2 = 0.
\end{equation}
According to the definition of $b$, only the positive solution of this
equation has physical meaning. Taking this into account, we obtain
that the value of $b$ is given by Eq. (\ref{b}). 

From the solution of Eq. (\ref{aaa}) one can see that the shock
reaches point $x$ at the moment $t_i(x) = x^2/b$. Right after the
shocks have passed the new value of $u$ at the moment $t_i(x)$ is
given by Eq. (\ref{um}). Therefore, according to Eq. (\ref{rest}) the
distribution $u'$ at a point $x$ and time $t > t_i(x)$ behind the
shock is given by
\begin{equation} \label{zz}
u'(x, t) = u_-[u\bigl(x, t_i(x)\bigr) ] + v t - v t_i(x),
\end{equation}
where $u$ is evaluated right before the shock reached $x$.  Since $u$
in Eq. (\ref{zz}) is close to 1, we can Taylor expand the dependence
$u_-(u_+)$ in this equation and keep only the first two terms. As a
result, we obtain
\begin{equation} \label{aaaa}
u'(x, t) \simeq u_-(1) + \kappa \left( a x^2 - {v x^2 \over b} \right)
- {v x^2 \over b} + v t,
\end{equation}
where $\kappa = - d u_- / d u_+$ evaluated at $u_+ = 1$. Using the
explicit expression for $u_-$ together with Eq. (\ref{cc}) in
Eq. (\ref{um}), one arrives at Eq. (\ref{kappa}). Then, using the
value of $b$ from Eq. (\ref{b}), Eq. (\ref{aaaa}) can be transformed
to Eq. (\ref{up}).

\bibliography{../main}

\end{document}